\documentclass[manuscript,screen]{acmart}
\usepackage{caption}
\usepackage{subcaption}
\usepackage{multirow}
\usepackage{lscape}
\usepackage{booktabs}
\usepackage[colorinlistoftodos, textsize=small, obeyFinal]{todonotes}

\AtBeginDocument{%
  }

\makeatletter
\let\@authorsaddresses\@empty
\makeatother

\begin{document}

\title{Autonomy for Older Adult-Agent Interaction}

\author{Jiaxin An}
\email{jiaxin.an@utexas.edu}
\orcid{0000-0003-2793-6469}
\affiliation{%
  \institution{The University of Texas at Austin}
  \city{Austin}
  \state{Texas}
  \country{USA}
}

\begin{abstract}
As the global population ages, artificial intelligence (AI)-powered agents have emerged as potential tools to support older adults' caregiving. Prior research has explored agent autonomy by identifying key interaction stages in task processes and defining the agent’s role at each stage. However, ensuring that agents align with older adults’ autonomy preferences remains a critical challenge. Drawing on interdisciplinary conceptualizations of autonomy, this paper examines four key dimensions of autonomy for older adults: decision-making autonomy, goal-oriented autonomy, control autonomy, and social responsibility autonomy. This paper then proposes the following research directions: (1) Addressing social responsibility autonomy, which concerns the ethical and social implications of agent use in communal settings; (2) Operationalizing agent autonomy from the task perspective; and (3) Developing autonomy measures. 
\end{abstract}


\maketitle

\section{Introduction}
By 2050, older adults are expected to comprise 16.0\% of the global population, placing considerable strain on healthcare systems to accommodate the needs of an aging society \cite{united_nations_department_of_economic_and_social_affairs_population_division_world_2020}. In response to this challenge, researchers have explored the deployment of artificial intelligence (AI)-powered agents to support various aspects of older adults’ daily lives. These applications encompass assistance with routine tasks, such as meal preparation \cite{czaja_designing_2019}, time management, and home maintenance \cite{even_benefits_2022, czaja_designing_2019}, as well as healthcare-related functions, including medication adherence, preventive care, and disease management \cite{czaja_designing_2019, even_benefits_2022}. Recent advancements in generative AI, particularly large language models (LLMs), have further expanded the capabilities of such agents. Unlike traditional rule-based or pre-programmed systems, LLM-powered agents can generate more nuanced and contextually appropriate responses, thereby improving user engagement and adaptability in human-agent interactions.

Although growing evidence highlights the positive impact of AI agents on older adults’ physical health (e.g, maintaining a healthy diet \cite{mccloud_using_2022}) and mental well-being (e.g., significantly decreased anxiety \cite{ryu_simple_2020}), challenges persist in ensuring these systems align with human values. This position paper focuses on the concept of autonomy, emphasizing its significance in older adult-agent interactions from two perspectives: (1) as a defining characteristic of an agent-based system and (2) as a fundamental requirement for older adults, especially when applying agents to support older adults' health and well-being. The paper concludes by outlining possible directions to build the alignment of agent autonomy and older adults' perceived autonomy, including several of my ongoing works.

\section{Autonomy for Machines: a Key Characteristic of An Agent}\label{sec: autonomy for machines}
The definition of an "agent" has long been debated, with new questions emerging as technologies continue to evolve rapidly \citep[p. 286]{bradshaw_humanagent_2011}. A central focus in agent-related research is \textit{the extent to which a system can operate independently in complex situations without continuous human supervision}. For instance, \citet{lewis_designing_1998} characterized a system as an agent if it could \textit{automate certain stages of human interaction}, such as anticipating user commands. To further explore levels of autonomy in agents, \citet{lewis_designing_1998} drew upon Norman’s twin gulfs of execution and evaluation \citep[p. 41-42]{norman_cognitive_1986}, which define seven stages of human-computer interaction:
(1) Establishing a goal;
(2) Forming an intention; 
(3) Specifying an action plan; 
(4) Executing the action; 
(5) Perceiving the system state; 
(6) Interpreting the system state; 
(7) Evaluating the outcome relative to the goal. 
Based on these stages, \citet{lewis_designing_1998} identified three distinct types of agents:
\begin{itemize}
    \item Semi-autonomous Agent: This agent automates all user activities except for goal setting and intention formation. An example is a price-comparison system that searches multiple databases and returns favorable results after users input their preferences. Semi-autonomous agents require users to have a detailed mental model of the system and interfaces that allow monitoring of task performance.
    \item Filter Agent: This agent automates the stages of perceiving, interpreting, and evaluating system states. A typical example is an email filter that assesses the importance of emails and presents the most relevant ones to users.
    \item Adaptive Agent: This agent automates the stages of intention, planning, and execution by learning from users' past actions and predicting future intentions to act accordingly.
\end{itemize}

While \citet{lewis_designing_1998} categorized agents based on their ability to automate different stages of human-computer interaction, \citet{cimolino_two_2022} examined autonomy through the lens of user control in task execution. They identified five roles that AI can assume in human-AI collaboration: 
(1) \textit{Supportive AI}: Assists users with specific tasks while keeping them in primary control. 
(2) \textit{Delegated AI}: Takes over specific tasks from the user. 
(3) \textit{Cooptable AI}: Performs certain tasks but remains under the user’s control. 
(4) \textit{Reciprocal AI}: Supports a dynamic relationship where tasks are delegated and reassigned between the user and the AI.
(5) \textit{Complementary AI}: Fully handles specific portions of a task, leaving others to the user. 

In summary, prior research identifies autonomy, or the degree of automation in task execution, as a defining characteristic of agents. Scholars have explored autonomy by analyzing both how agents automate stages of human-computer interaction and the extent of user control in task execution. This perspective highlights the need to:
(1) Identify key interaction stages in specific task processes.
(2) Define the role of agents at each stage, determining the appropriate balance between user control and system autonomy.

\section{Autonomy for Older Adults: A Fundamental Need}
Unlike the autonomy of an agent, which primarily refers to its capacity to function independently, perceived autonomy in humans is a more complex and multidimensional concept. \citet{kim_exploring_2024} conducted a review of how autonomy is conceptualized across various disciplines, including philosophy, psychology, medical ethics, and feminism, highlighting key distinctions, as summarized in Table \ref{tab:autonomy}:

\begin{table*}[ht]
    \centering
    \setlength{\arrayrulewidth}{0mm}
    \begin{tabular}{l|p{3cm}|p{9cm}}
    \toprule
       Domain  & Summary & Argument examples\\
    \midrule
       Philosophy  & Moral value & Self-governance of city status \cite{saad_history_2018}, self-legislation in a social contract \cite{rousseau_rousseau_2018}, the basis for human dignity that one's decision is guided by the maximum of internal rationality \cite{kant_kant_2017}\\
       Psychology  & Internally originating sense of self control & A type of internal motivation in self-determination theory \cite{ryan_self-determination_2000}, locus of control and perceived control \cite{rotter_generalized_1966} \\
       Medical ethics  &  Patients can make independent decisions & Be free from the controlling influence of others, limitations that prevent meaningful choices, and inadequate understanding \cite{childress_principles_1994} \\
       Feminism  & Relational autonomy & Instead of acting independently, an agent can still exercise autonomy by delegating the tasks to those who can execute the agent's decision \cite{collopy_autonomy_1988}. \\
    \bottomrule
    \end{tabular}
    \caption{The conceptualization of autonomy in different disciplines}
    \label{tab:autonomy}
\end{table*}

Building on this review, \citet{kim_exploring_2024} proposed a framework outlining four dimensions of autonomy:
\begin{itemize}
    \item Power to Make Decisions – The freedom to make informed choices without external coercion.
    \item Capacity for Action – The ability to utilize available resources to achieve self-directed goals.
    \item Sense of Control – The perception that one can influence their environment and life.
    \item Self-Governance – The ability to manage one’s actions responsibly within a social context.
\end{itemize}
Compared to younger individuals, older adults face distinct challenges in maintaining autonomy across these dimensions due to age-related factors, such as: 
(1) Physical decline. Aging leads to unavoidable deterioration in vision, hearing, immune function, and overall physiological capacity (e.g., reduced respiratory function and cardiac efficiency) \cite{papalia_human_2009, cavanaugh_adult_2019}. These changes can significantly affect older adults' capacity for action and sense of control. 
(2) Cognitive decline. Older adults experience declines in fundamental cognitive processes, such as information processing and problem-solving \citep[p. 188]{cavanaugh_adult_2019}, which may undermine their power to make decisions.
(3) Ageism. Negative stereotypes (e.g., the belief that older adults are weak or incapable of contributing to society), prejudices, and discriminatory attitudes can have profound psychological consequences. Internalized ageism has been shown to diminish older adults' self-perception, elevate the risk of mental health issues \cite{reynolds_mental_2022}, and increase susceptibility to suicidal ideation \cite{gendron_internalized_2024}.

To counteract these challenges and preserve their autonomy, older adults are increasingly open to adopting AI agents \cite{santini_digital_2020, santini_user_2021}. However, they express strong concerns about over-reliance on such systems and emphasize the importance of maintaining control in their interactions, especially in health contexts. For instance, \citet{santini_user_2021} found that many older adults preferred reactive agents, allowing them to retain full decision-making authority over whether to engage with a system’s functions. While some appreciated proactive features, such as exercise reminders, they insisted on having the option to disable them at will \cite{tsiourti_virtual_2014}. Additionally, older adults desired control over agent operations, including the ability to power off the system as needed \cite{vardoulakis_designing_2012}, as well as oversight over the sharing of personal data, particularly with healthcare providers. \citet{brewer_if_2022} further noted that older adults preferred agents to communicate information in ways that aligned with their self-image. For example, one participant expressed a desire for the agent to portray them as a positive person, stating, “\textit{I’d like agents to tell whoever that I’m a happy person}.” These concerns and preferences closely align with broader conceptions of autonomy, emphasizing older adults’ need to make independent decisions in agent interactions, utilize AI agents to support personal goals, retain control over system operations, and manage their engagement with agents in a socially responsible manner \cite{kim_exploring_2024}.

\section{Conclusion and Future Research}
This paper reviewed autonomy on both sides: the task automation capacity of an agent and the nuanced but fundamental value for older adults, especially when applying agents to support older adults' health and well-being. As the alignment of agent autonomy and older adults' perceived autonomy is essential but understudied, this section aims to propose several possible directions to address the gap.

\subsection{Social responsibility autonomy: an understudied area}
In my recent systematic literature review (under review), I examined how existing studies about "conversational agents (CAs) for older adults' health" have addressed the four dimensions of autonomy \cite{kim_exploring_2021}. These dimensions include:
\begin{itemize}
    \item Decision-making autonomy: Allowing older adults to make informed choices when interacting with CAs.
    \item Goal-oriented autonomy: Enabling older adults to use CAs as tools to achieve specific health-related or lifestyle objectives.
    \item Control autonomy: Providing mechanisms for older adults to manage and configure CAs according to their preferences.
    \item Social responsibility autonomy: Supporting older adults in responsibly integrating CAs into their social environments and ethical frameworks. 
\end{itemize}

As summarized in Table \ref{tab:autonomy example study}, while prior research has explored the first three dimensions—decision-making, goal-oriented, and control autonomy—there remains a significant gap concerning social responsibility autonomy. This dimension extends beyond individual autonomy to consider the broader social and ethical implications of older adults’ interactions with CAs. For instance, it encompasses how older adults balance virtual and face-to-face communication when using CAs and how these systems integrate responsibly into communal settings such as nursing homes.

\begin{table*}[ht]
    \centering
    \setlength{\arrayrulewidth}{0mm}
    \begin{tabular}{p{4cm}|p{10cm}}
    \toprule
       Dimensions of autonomy  & Example studies\\
    \midrule
       Decision-making autonomy  & A Wizard-of-Oz study found that low-income older adults expected CAs to assist with health-related decision-making based on verified information, such as recommending medication or advising whether to visit a doctor \cite{nallam_question_2020}. Given the high cost of medical consultations, older adults in this study were willing to accept CA recommendations, provided the system explicitly indicated that the advice was based on validated sources.\\
       Goal-oriented autonomy  & Older adults expressed interest in using CAs as a resource for retirement planning. For example, they envisioned leveraging CAs to gather financial information needed for maintaining a healthy and comfortable post-retirement life \cite{santini_digital_2020}.\\
       Control autonomy  &  Older adults emphasized the importance of maintaining full control over CAs, including: (1) Adjusting proactive features, such as exercise reminders, to avoid feeling monitored or pressured by the system \cite{santini_digital_2020}; (2) Managing data recording and sharing processes to protect privacy \cite{tsiourti_virtual_2014} and ensure that their self-presentation aligns with their desired public image \cite{brewer_empirical_2022}. \\
       Social responsibility autonomy  & No existing studies explicitly address this dimension.  \\
    \bottomrule
    \end{tabular}
    \caption{Existing research on autonomy dimensions in older adults’ interactions with CAs}
    \label{tab:autonomy example study}
\end{table*}

This gap highlights the need for future research to explore how older adults can navigate the societal and ethical complexities of agent integration. Key questions include how agents might shape social relationships, influence older adults’ perceptions of social engagement, and function responsibly within shared living spaces. Investigating these aspects will provide a more comprehensive understanding of autonomy in human-agent interaction, ensuring that agents support not only individual decision-making and goal achievement but also the broader social and ethical considerations of aging populations.

\subsection{Operationalize agent autonomy from the task perspective}
Prior research has emphasized the importance of designing agents by identifying key interaction stages within specific task processes and defining the agent’s role at each stage (see Section \ref{sec: autonomy for machines}). Given that agents can be applied across diverse healthcare scenarios, there remains a substantial opportunity for future research to specify the task processes involved in each scenario, determine the levels of autonomy agents can achieve, and identify the optimal level of autonomy based on older adults’ preferences.

Table \ref{tab:autonomy in health info seeking} illustrates a potential framework for designing agents that support varying levels of decision-making autonomy in different stages of the health information-seeking process. An agent with low decision-making autonomy relies heavily on older adults to clarify their needs, select information sources, digest information, and make decisions. Conversely, an agent with high decision-making autonomy adopts a more proactive role, assisting users by anticipating their needs, generating search queries, synthesizing relevant information, and recommending search refinements. While higher autonomy can reduce the effort required for health-related research, it also raises concerns about potential over-reliance on AI agents. To determine the appropriate level of autonomy, a systematic evaluation of older adults’ perceived autonomy is necessary, which will be discussed in the following section.

\begin{table*}[ht]
    \centering
    \setlength{\arrayrulewidth}{0mm}
    \begin{tabular}{p{5cm}|p{9cm}}
    \toprule
       Health information seeking stages  & Levels of agent decision-making autonomy (from low to high)\\
    \midrule
       Identify health information needs  & Passively receive older adults' requests -> Proactively recommend health information that older adults might be interested in\\
       Select information sources  & Search on sources determined by older adults -> Determine the source of searches based on older adults' past search history.\\
       Question formulation & Passively receive older adults' questions -> Generate search queries based on older adults' description \\
       Browse and digest information  &  Provide the top search results to older adults -> Synthesize information based on older adults' request -> Synthesize information and highlight the part that older adults would be interested in.\\
       Update information needs & Passively receive older adults' requests -> Proactively suggest older adults modify search direction.\\
    \bottomrule
    \end{tabular}
    \caption{Possible levels of agent autonomy in stages of health information seeking}
    \label{tab:autonomy in health info seeking}
\end{table*}
%As researchers are imagining the level of automation that an agent can be, they also noticed the ethical concerns. \citet{} pointed out two important trade-off for human-agent interaction: (1) the degree to which the agent's automation needs users' initiation. (2) the degree to which the agent provide explicit feedback to users. As individuals want to delegate tasks to agents, they still need different levels of control, i.e., initiating agents and get their feedback, depending on the contexts. 

\subsection{Develop Autonomy Measures}
To systematically assess older adults’ perceptions of autonomy, I plan to adapt existing scales from psychology to fit the context of older adults' agent use. Currently, I am reviewing established autonomy measurement scales, such as those proposed by \citet{antsaklis_autonomy_2020}, \citet{bekker_short_2006}, and \citet{breaugh_measurement_1985}. These scales, originally developed to measure general perceptions of autonomy in decision-making and behavioral contexts, may require modifications to address specific aspects of human-agent interactions. Potential adaptations include adding items that assess:
\begin{itemize}
    \item Decision-making autonomy: Do users feel they can direct the flow of the conversation with the agent?
    \item Control autonomy: Can users modify the agent settings and responses to match their preferences?
    \item Goal-oriented autonomy: Do users feel supported in achieving their health-related goals through the agent's assistance?
    \item Social responsibility autonomy: Are users aware of and comfortable with the broader social and ethical implications of using the agent?
\end{itemize}

\bibliographystyle{ACM-Reference-Format}
\bibliography{references_group.bib, additional.bib}


\begin{thebibliography}{32}

%%% the Web address.

\ifx \showCODEN    \undefined \def \showCODEN     #1{\unskip}     \fi
\ifx \showDOI      \undefined \def \showDOI       #1{#1}\fi
\ifx \showISBNx    \undefined \def \showISBNx     #1{\unskip}     \fi
\ifx \showISBNxiii \undefined \def \showISBNxiii  #1{\unskip}     \fi
\ifx \showISSN     \undefined \def \showISSN      #1{\unskip}     \fi
\ifx \showLCCN     \undefined \def \showLCCN      #1{\unskip}     \fi
\ifx \shownote     \undefined \def \shownote      #1{#1}          \fi
\ifx \showarticletitle \undefined \def \showarticletitle #1{#1}   \fi
\ifx \showURL      \undefined \def \showURL       {\relax}        \fi
% invisible to TeX
\providecommand\bibfield[2]{#2}
\providecommand\bibinfo[2]{#2}
\providecommand\natexlab[1]{#1}
\providecommand\showeprint[2][]{arXiv:#2}

\bibitem[Antsaklis(2020)]%
        {antsaklis_autonomy_2020}
\bibfield{author}{\bibinfo{person}{Panos Antsaklis}.} \bibinfo{year}{2020}\natexlab{}.
\newblock \showarticletitle{Autonomy and metrics of autonomy}.
\newblock \bibinfo{journal}{\emph{Annual Reviews in Control}}  \bibinfo{volume}{49} (\bibinfo{date}{Jan.} \bibinfo{year}{2020}), \bibinfo{pages}{15--26}.
\newblock
\showISSN{1367-5788}
\urldef\tempurl%
\url{https://doi.org/10.1016/j.arcontrol.2020.05.001}
\showDOI{\tempurl}


\bibitem[Bekker and van Assen(2006)]%
        {bekker_short_2006}
\bibfield{author}{\bibinfo{person}{Marrie H.~J. Bekker} {and} \bibinfo{person}{Marcel A. L.~M. van Assen}.} \bibinfo{year}{2006}\natexlab{}.
\newblock \showarticletitle{A {Short} {Form} of the {Autonomy} {Scale}: {Properties} of the {Autonomy}–{Connectedness} {Scale} ({ACS}–30)}.
\newblock \bibinfo{journal}{\emph{Journal of Personality Assessment}} \bibinfo{volume}{86}, \bibinfo{number}{1} (\bibinfo{date}{Feb.} \bibinfo{year}{2006}), \bibinfo{pages}{51--60}.
\newblock
\showISSN{0022-3891}
\urldef\tempurl%
\url{https://doi.org/10.1207/s15327752jpa8601_07}
\showDOI{\tempurl}
\newblock
\shownote{Publisher: Routledge \_eprint: https://doi.org/10.1207/s15327752jpa8601\_07}.


\bibitem[Bradshaw et~al\mbox{.}(2011)]%
        {bradshaw_humanagent_2011}
\bibfield{author}{\bibinfo{person}{Jeffrey~M. Bradshaw}, \bibinfo{person}{Paul~J. Feltovich}, {and} \bibinfo{person}{Matthew Johnson}.} \bibinfo{year}{2011}\natexlab{}.
\newblock \showarticletitle{Human–{Agent} {Interaction}}.
\newblock In \bibinfo{booktitle}{\emph{The {Handbook} of {Human}-{Machine} {Interaction}}}. \bibinfo{publisher}{CRC Press}.
\newblock
\showISBNx{978-1-315-55738-0}
\newblock
\shownote{Num Pages: 18}.


\bibitem[Breaugh(1985)]%
        {breaugh_measurement_1985}
\bibfield{author}{\bibinfo{person}{James~A. Breaugh}.} \bibinfo{year}{1985}\natexlab{}.
\newblock \showarticletitle{The {Measurement} of {Work} {Autonomy}}.
\newblock \bibinfo{journal}{\emph{Human Relations}} \bibinfo{volume}{38}, \bibinfo{number}{6} (\bibinfo{date}{June} \bibinfo{year}{1985}), \bibinfo{pages}{551--570}.
\newblock
\showISSN{0018-7267}
\urldef\tempurl%
\url{https://doi.org/10.1177/001872678503800604}
\showDOI{\tempurl}
\newblock
\shownote{Publisher: SAGE Publications Ltd}.


\bibitem[Brewer et~al\mbox{.}(2022)]%
        {brewer_empirical_2022}
\bibfield{author}{\bibinfo{person}{Robin Brewer}, \bibinfo{person}{Casey Pierce}, \bibinfo{person}{Pooja Upadhyay}, {and} \bibinfo{person}{Leeseul Park}.} \bibinfo{year}{2022}\natexlab{}.
\newblock \showarticletitle{An {Empirical} {Study} of {Older} {Adult}’s {Voice} {Assistant} {Use} for {Health} {Information} {Seeking}}.
\newblock \bibinfo{journal}{\emph{ACM Transactions on Interactive Intelligent Systems}} \bibinfo{volume}{12}, \bibinfo{number}{2} (\bibinfo{date}{June} \bibinfo{year}{2022}), \bibinfo{pages}{1--32}.
\newblock
\showISSN{2160-6455, 2160-6463}
\urldef\tempurl%
\url{https://doi.org/10.1145/3484507}
\showDOI{\tempurl}


\bibitem[Brewer(2022)]%
        {brewer_if_2022}
\bibfield{author}{\bibinfo{person}{Robin~N Brewer}.} \bibinfo{year}{2022}\natexlab{}.
\newblock \showarticletitle{“{If} {Alexa} knew the state {I} was in, it would cry”: {Older} {Adults}’ {Perspectives} of {Voice} {Assistants} for {Health}}. \bibinfo{pages}{8}.
\newblock


\bibitem[Cavanaugh and Blanchard-Fields(2019)]%
        {cavanaugh_adult_2019}
\bibfield{author}{\bibinfo{person}{John~C. Cavanaugh} {and} \bibinfo{person}{Fredda Blanchard-Fields}.} \bibinfo{year}{2019}\natexlab{}.
\newblock \bibinfo{booktitle}{\emph{Adult development and aging} (\bibinfo{edition}{eighth edition} ed.)}.
\newblock \bibinfo{publisher}{Cengage Learning}, \bibinfo{address}{Boston, MA}.
\newblock
\showISBNx{978-1-337-55908-9 978-1-337-56352-9}


\bibitem[Childress and Beauchamp(1994)]%
        {childress_principles_1994}
\bibfield{author}{\bibinfo{person}{James~F Childress} {and} \bibinfo{person}{Tom~L Beauchamp}.} \bibinfo{year}{1994}\natexlab{}.
\newblock \bibinfo{booktitle}{\emph{Principles of biomedical ethics}}.
\newblock \bibinfo{publisher}{Oxford University Press Oxford.–}.
\newblock


\bibitem[Cimolino and Graham(2022)]%
        {cimolino_two_2022}
\bibfield{author}{\bibinfo{person}{Gabriele Cimolino} {and} \bibinfo{person}{T.C.~Nicholas Graham}.} \bibinfo{year}{2022}\natexlab{}.
\newblock \showarticletitle{Two {Heads} {Are} {Better} {Than} {One}: {A} {Dimension} {Space} for {Unifying} {Human} and {Artificial} {Intelligence} in {Shared} {Control}}. In \bibinfo{booktitle}{\emph{{CHI} {Conference} on {Human} {Factors} in {Computing} {Systems}}}. \bibinfo{publisher}{ACM}, \bibinfo{address}{New Orleans LA USA}, \bibinfo{pages}{1--21}.
\newblock
\showISBNx{978-1-4503-9157-3}
\urldef\tempurl%
\url{https://doi.org/10.1145/3491102.3517610}
\showDOI{\tempurl}


\bibitem[Collopy(1988)]%
        {collopy_autonomy_1988}
\bibfield{author}{\bibinfo{person}{Bart~J. Collopy}.} \bibinfo{year}{1988}\natexlab{}.
\newblock \showarticletitle{Autonomy in {Long} {Term} {Care}: {Some} {Crucial} {Distinctions1}}.
\newblock \bibinfo{journal}{\emph{The Gerontologist}} \bibinfo{volume}{28}, \bibinfo{number}{Suppl} (\bibinfo{date}{June} \bibinfo{year}{1988}), \bibinfo{pages}{10--17}.
\newblock
\showISSN{0016-9013}
\urldef\tempurl%
\url{https://doi.org/10.1093/geront/28.Suppl.10}
\showDOI{\tempurl}


\bibitem[Czaja(2019)]%
        {czaja_designing_2019}
\bibfield{author}{\bibinfo{person}{Sara~J Czaja}.} \bibinfo{year}{2019}\natexlab{}.
\newblock \bibinfo{booktitle}{\emph{Designing for {Older} {Adults}: {Principles} and {Creative} {Human} {Factors} {Approaches}}}.
\newblock \bibinfo{publisher}{CRC press}.
\newblock


\bibitem[Even et~al\mbox{.}(2022)]%
        {even_benefits_2022}
\bibfield{author}{\bibinfo{person}{Christiane Even}, \bibinfo{person}{Torsten Hammann}, \bibinfo{person}{Vera Heyl}, \bibinfo{person}{Christian Rietz}, \bibinfo{person}{Hans-Werner Wahl}, \bibinfo{person}{Peter Zentel}, {and} \bibinfo{person}{Anna Schlomann}.} \bibinfo{year}{2022}\natexlab{}.
\newblock \showarticletitle{Benefits and challenges of conversational agents in older adults: {A} scoping review}.
\newblock \bibinfo{journal}{\emph{Zeitschrift für Gerontologie und Geriatrie}} (\bibinfo{date}{July} \bibinfo{year}{2022}).
\newblock
\showISSN{1435-1269}
\urldef\tempurl%
\url{https://doi.org/10.1007/s00391-022-02085-9}
\showDOI{\tempurl}


\bibitem[Gendron et~al\mbox{.}(2024)]%
        {gendron_internalized_2024}
\bibfield{author}{\bibinfo{person}{Tracey Gendron}, \bibinfo{person}{Alyssa Camp}, \bibinfo{person}{Gigi Amateau}, {and} \bibinfo{person}{Kanako Iwanaga}.} \bibinfo{year}{2024}\natexlab{}.
\newblock \showarticletitle{Internalized ageism as a risk factor for suicidal ideation in later life}.
\newblock \bibinfo{journal}{\emph{Aging \& Mental Health}} \bibinfo{volume}{28}, \bibinfo{number}{4} (\bibinfo{date}{April} \bibinfo{year}{2024}), \bibinfo{pages}{701--705}.
\newblock
\showISSN{1360-7863}
\urldef\tempurl%
\url{https://doi.org/10.1080/13607863.2023.2271870}
\showDOI{\tempurl}
\newblock
\shownote{Publisher: Routledge \_eprint: https://doi.org/10.1080/13607863.2023.2271870}.


\bibitem[Kant(2017)]%
        {kant_kant_2017}
\bibfield{author}{\bibinfo{person}{Immanuel Kant}.} \bibinfo{year}{2017}\natexlab{}.
\newblock \bibinfo{booktitle}{\emph{Kant: {The} {Metaphysics} of {Morals}}}.
\newblock \bibinfo{publisher}{Cambridge University Press}.
\newblock
\showISBNx{978-1-107-08639-5}
\newblock
\shownote{Google-Books-ID: 3hk0DwAAQBAJ}.


\bibitem[Kim(2024)]%
        {kim_exploring_2024}
\bibfield{author}{\bibinfo{person}{Miso Kim}.} \bibinfo{year}{2024}\natexlab{}.
\newblock \showarticletitle{Exploring {Autonomy} as a {Design} {Principle}: {Theoretical} {Review} of {Autonomy} and {Case} {Studies} of {Service} {Design} for {Seniors}}.
\newblock \bibinfo{journal}{\emph{Design and Culture}} \bibinfo{volume}{16}, \bibinfo{number}{1} (\bibinfo{date}{Jan.} \bibinfo{year}{2024}), \bibinfo{pages}{83--107}.
\newblock
\showISSN{1754-7075, 1754-7083}
\urldef\tempurl%
\url{https://doi.org/10.1080/17547075.2023.2259658}
\showDOI{\tempurl}


\bibitem[Kim and Choudhury(2021)]%
        {kim_exploring_2021}
\bibfield{author}{\bibinfo{person}{Sunyoung Kim} {and} \bibinfo{person}{Abhishek Choudhury}.} \bibinfo{year}{2021}\natexlab{}.
\newblock \showarticletitle{Exploring older adults’ perception and use of smart speaker-based voice assistants: {A} longitudinal study}.
\newblock \bibinfo{journal}{\emph{Computers in Human Behavior}}  \bibinfo{volume}{124} (\bibinfo{date}{Nov.} \bibinfo{year}{2021}), \bibinfo{pages}{106914}.
\newblock
\showISSN{0747-5632}
\urldef\tempurl%
\url{https://doi.org/10.1016/j.chb.2021.106914}
\showDOI{\tempurl}


\bibitem[Lewis(1998)]%
        {lewis_designing_1998}
\bibfield{author}{\bibinfo{person}{Michael Lewis}.} \bibinfo{year}{1998}\natexlab{}.
\newblock \showarticletitle{Designing for {Human}-{Agent} {Interaction}}.
\newblock \bibinfo{journal}{\emph{AI Magazine}} \bibinfo{volume}{19}, \bibinfo{number}{2} (\bibinfo{year}{1998}), \bibinfo{pages}{67--78}.
\newblock


\bibitem[McCloud et~al\mbox{.}(2022)]%
        {mccloud_using_2022}
\bibfield{author}{\bibinfo{person}{Rachel McCloud}, \bibinfo{person}{Carly Perez}, \bibinfo{person}{Mesfin~Awoke Bekalu}, {and} \bibinfo{person}{K. Viswanath}.} \bibinfo{year}{2022}\natexlab{}.
\newblock \showarticletitle{Using {Smart} {Speaker} {Technology} for {Health} and {Well}-being in an {Older} {Adult} {Population}: {Pre}-{Post} {Feasibility} {Study}}.
\newblock \bibinfo{journal}{\emph{JMIR Aging}} \bibinfo{volume}{5}, \bibinfo{number}{2} (\bibinfo{date}{May} \bibinfo{year}{2022}), \bibinfo{pages}{e33498}.
\newblock
\urldef\tempurl%
\url{https://doi.org/10.2196/33498}
\showDOI{\tempurl}
\newblock
\shownote{Company: JMIR Aging Distributor: JMIR Aging Institution: JMIR Aging Label: JMIR Aging Publisher: JMIR Publications Inc., Toronto, Canada}.


\bibitem[Nallam et~al\mbox{.}(2020)]%
        {nallam_question_2020}
\bibfield{author}{\bibinfo{person}{Phani Nallam}, \bibinfo{person}{Siddhant Bhandari}, \bibinfo{person}{Jamie Sanders}, {and} \bibinfo{person}{Aqueasha Martin-Hammond}.} \bibinfo{year}{2020}\natexlab{}.
\newblock \showarticletitle{A {Question} of {Access}: {Exploring} the {Perceived} {Benefits} and {Barriers} of {Intelligent} {Voice} {Assistants} for {Improving} {Access} to {Consumer} {Health} {Resources} {Among} {Low}-{Income} {Older} {Adults}}.
\newblock \bibinfo{journal}{\emph{Gerontology and Geriatric Medicine}}  \bibinfo{volume}{6} (\bibinfo{date}{Jan.} \bibinfo{year}{2020}), \bibinfo{pages}{2333721420985975}.
\newblock
\showISSN{2333-7214}
\urldef\tempurl%
\url{https://doi.org/10.1177/2333721420985975}
\showDOI{\tempurl}
\newblock
\shownote{Publisher: SAGE Publications Inc}.


\bibitem[Norman(1986)]%
        {norman_cognitive_1986}
\bibfield{author}{\bibinfo{person}{Donald~Arthur Norman}.} \bibinfo{year}{1986}\natexlab{}.
\newblock \showarticletitle{Cognitive {Engineering}}.
\newblock In \bibinfo{booktitle}{\emph{User {Centered} {System} {Design}: {New} {Perspectives} on {Human}-{Computer} {Interaction}}}. \bibinfo{pages}{31--61}.
\newblock
\showISBNx{978-0-367-80732-0}
\urldef\tempurl%
\url{https://doi.org/10.1201/b15703-3}
\showDOI{\tempurl}
\newblock
\shownote{Journal Abbreviation: User Centered System Design: New Perspectives on Human-Computer Interaction}.


\bibitem[Papalia et~al\mbox{.}(2009)]%
        {papalia_human_2009}
\bibfield{author}{\bibinfo{person}{Diane~E. Papalia}, \bibinfo{person}{Sally~Wendkos Olds}, {and} \bibinfo{person}{Ruth~Duskin Feldman}.} \bibinfo{year}{2009}\natexlab{}.
\newblock \bibinfo{booktitle}{\emph{Human development} (\bibinfo{edition}{eleventh edition} ed.)}.
\newblock \bibinfo{publisher}{McGraw-Hill}, \bibinfo{address}{New York}.
\newblock
\showISBNx{978-0-07-337016-3}
\newblock
\shownote{OCLC: 226356755}.


\bibitem[Reynolds et~al\mbox{.}(2022)]%
        {reynolds_mental_2022}
\bibfield{author}{\bibinfo{person}{Charles~F. Reynolds}, \bibinfo{person}{Dilip~V. Jeste}, \bibinfo{person}{Perminder~S. Sachdev}, {and} \bibinfo{person}{Dan~G. Blazer}.} \bibinfo{year}{2022}\natexlab{}.
\newblock \showarticletitle{Mental health care for older adults: recent advances and new directions in clinical practice and research}.
\newblock \bibinfo{journal}{\emph{World Psychiatry}} \bibinfo{volume}{21}, \bibinfo{number}{3} (\bibinfo{date}{Oct.} \bibinfo{year}{2022}), \bibinfo{pages}{336--363}.
\newblock
\showISSN{1723-8617, 2051-5545}
\urldef\tempurl%
\url{https://doi.org/10.1002/wps.20996}
\showDOI{\tempurl}


\bibitem[Rotter(1966)]%
        {rotter_generalized_1966}
\bibfield{author}{\bibinfo{person}{Julian~B. Rotter}.} \bibinfo{year}{1966}\natexlab{}.
\newblock \showarticletitle{Generalized expectancies for internal versus external control of reinforcement}.
\newblock \bibinfo{journal}{\emph{Psychological Monographs: General and Applied}} \bibinfo{volume}{80}, \bibinfo{number}{1} (\bibinfo{year}{1966}), \bibinfo{pages}{1--28}.
\newblock
\showISSN{0096-9753}
\urldef\tempurl%
\url{https://doi.org/10.1037/h0092976}
\showDOI{\tempurl}
\newblock
\shownote{Place: US Publisher: American Psychological Association}.


\bibitem[Rousseau(2018)]%
        {rousseau_rousseau_2018}
\bibfield{author}{\bibinfo{person}{Jean-Jacques Rousseau}.} \bibinfo{year}{2018}\natexlab{}.
\newblock \bibinfo{booktitle}{\emph{Rousseau: {The} {Social} {Contract} and {Other} {Later} {Political} {Writings}}}.
\newblock \bibinfo{publisher}{Cambridge University Press}.
\newblock
\showISBNx{978-1-108-57857-8}
\newblock
\shownote{Google-Books-ID: N3g6EAAAQBAJ}.


\bibitem[Ryan and Deci(2000)]%
        {ryan_self-determination_2000}
\bibfield{author}{\bibinfo{person}{Richard~M. Ryan} {and} \bibinfo{person}{Edward~L. Deci}.} \bibinfo{year}{2000}\natexlab{}.
\newblock \showarticletitle{Self-determination theory and the facilitation of intrinsic motivation, social development, and well-being}.
\newblock \bibinfo{journal}{\emph{American Psychologist}} \bibinfo{volume}{55}, \bibinfo{number}{1} (\bibinfo{year}{2000}), \bibinfo{pages}{68--78}.
\newblock
\showISSN{1935-990X}
\urldef\tempurl%
\url{https://doi.org/10.1037/0003-066X.55.1.68}
\showDOI{\tempurl}
\newblock
\shownote{Place: US Publisher: American Psychological Association}.


\bibitem[Ryu et~al\mbox{.}(2020)]%
        {ryu_simple_2020}
\bibfield{author}{\bibinfo{person}{Hyeyoung Ryu}, \bibinfo{person}{Soyeon Kim}, \bibinfo{person}{Dain Kim}, \bibinfo{person}{Sooan Han}, \bibinfo{person}{Keeheon Lee}, {and} \bibinfo{person}{Younah Kang}.} \bibinfo{year}{2020}\natexlab{}.
\newblock \showarticletitle{Simple and {Steady} {Interactions} {Win} the {Healthy} {Mentality}: {Designing} a {Chatbot} {Service} for the {Elderly}}.
\newblock \bibinfo{journal}{\emph{Proceedings of the ACM on Human-Computer Interaction}} \bibinfo{volume}{4}, \bibinfo{number}{CSCW2} (\bibinfo{date}{Oct.} \bibinfo{year}{2020}), \bibinfo{pages}{152:1--152:25}.
\newblock
\urldef\tempurl%
\url{https://doi.org/10.1145/3415223}
\showDOI{\tempurl}


\bibitem[Saad(2018)]%
        {saad_history_2018}
\bibfield{author}{\bibinfo{person}{Toni~C. Saad}.} \bibinfo{year}{2018}\natexlab{}.
\newblock \showarticletitle{The history of autonomy in medicine from antiquity to principlism}.
\newblock \bibinfo{journal}{\emph{Medicine, Health Care and Philosophy}} \bibinfo{volume}{21}, \bibinfo{number}{1} (\bibinfo{date}{March} \bibinfo{year}{2018}), \bibinfo{pages}{125--137}.
\newblock
\showISSN{1572-8633}
\urldef\tempurl%
\url{https://doi.org/10.1007/s11019-017-9781-2}
\showDOI{\tempurl}


\bibitem[Santini et~al\mbox{.}(2020)]%
        {santini_digital_2020}
\bibfield{author}{\bibinfo{person}{Sara Santini}, \bibinfo{person}{Flavia Galassi}, \bibinfo{person}{Johannes Kropf}, {and} \bibinfo{person}{Vera Stara}.} \bibinfo{year}{2020}\natexlab{}.
\newblock \showarticletitle{A {Digital} {Coach} {Promoting} {Healthy} {Aging} among {Older} {Adults} in {Transition} to {Retirement}: {Results} from a {Qualitative} {Study} in {Italy}}.
\newblock \bibinfo{journal}{\emph{Sustainability}} \bibinfo{volume}{12}, \bibinfo{number}{18} (\bibinfo{date}{Jan.} \bibinfo{year}{2020}), \bibinfo{pages}{7400}.
\newblock
\showISSN{2071-1050}
\urldef\tempurl%
\url{https://doi.org/10.3390/su12187400}
\showDOI{\tempurl}
\newblock
\shownote{Number: 18 Publisher: Multidisciplinary Digital Publishing Institute}.


\bibitem[Santini et~al\mbox{.}(2021)]%
        {santini_user_2021}
\bibfield{author}{\bibinfo{person}{Sara Santini}, \bibinfo{person}{Vera Stara}, \bibinfo{person}{Flavia Galassi}, \bibinfo{person}{Alessandra Merizzi}, \bibinfo{person}{Cornelia Schneider}, \bibinfo{person}{Sabine Schwammer}, \bibinfo{person}{Elske Stolte}, {and} \bibinfo{person}{Johannes Kropf}.} \bibinfo{year}{2021}\natexlab{}.
\newblock \showarticletitle{User {Requirements} {Analysis} of an {Embodied} {Conversational} {Agent} for {Coaching} {Older} {Adults} to {Choose} {Active} and {Healthy} {Ageing} {Behaviors} during the {Transition} to {Retirement}: {A} {Cross}-{National} {User} {Centered} {Design} {Study}}.
\newblock \bibinfo{journal}{\emph{International Journal of Environmental Research and Public Health}} \bibinfo{volume}{18}, \bibinfo{number}{18} (\bibinfo{date}{Jan.} \bibinfo{year}{2021}), \bibinfo{pages}{9681}.
\newblock
\showISSN{1660-4601}
\urldef\tempurl%
\url{https://doi.org/10.3390/ijerph18189681}
\showDOI{\tempurl}
\newblock
\shownote{Number: 18 Publisher: Multidisciplinary Digital Publishing Institute}.


\bibitem[Tsiourti et~al\mbox{.}(2014)]%
        {tsiourti_virtual_2014}
\bibfield{author}{\bibinfo{person}{Christiana Tsiourti}, \bibinfo{person}{Emilie Joly}, \bibinfo{person}{Cindy Wings}, \bibinfo{person}{Maher~Ben Moussa}, {and} \bibinfo{person}{Katarzyna Wac}.} \bibinfo{year}{2014}\natexlab{}.
\newblock \showarticletitle{Virtual assistive companions for older adults: qualitative field study and design implications}. In \bibinfo{booktitle}{\emph{Proceedings of the 8th {International} {Conference} on {Pervasive} {Computing} {Technologies} for {Healthcare}}} \emph{(\bibinfo{series}{{PervasiveHealth} '14})}. \bibinfo{publisher}{ICST (Institute for Computer Sciences, Social-Informatics and Telecommunications Engineering)}, \bibinfo{address}{Brussels, BEL}, \bibinfo{pages}{57--64}.
\newblock
\showISBNx{978-1-63190-011-2}
\urldef\tempurl%
\url{https://doi.org/10.4108/icst.pervasivehealth.2014.254943}
\showDOI{\tempurl}


\bibitem[{United Nations. Department of Economic and Social Affairs. Population Division}(2020)]%
        {united_nations_department_of_economic_and_social_affairs_population_division_world_2020}
\bibfield{author}{\bibinfo{person}{{United Nations. Department of Economic and Social Affairs. Population Division}}.} \bibinfo{year}{2020}\natexlab{}.
\newblock \bibinfo{booktitle}{\emph{World population ageing 2020 highlights: {Living} arrangements of older persons}}.
\newblock \bibinfo{publisher}{UN}.
\newblock
\showISBNx{92-1-005193-9}


\bibitem[Vardoulakis et~al\mbox{.}(2012)]%
        {vardoulakis_designing_2012}
\bibfield{author}{\bibinfo{person}{Laura~Pfeifer Vardoulakis}, \bibinfo{person}{Lazlo Ring}, \bibinfo{person}{Barbara Barry}, \bibinfo{person}{Candace~L. Sidner}, {and} \bibinfo{person}{Timothy Bickmore}.} \bibinfo{year}{2012}\natexlab{}.
\newblock \showarticletitle{Designing relational agents as long term social companions for older adults}. In \bibinfo{booktitle}{\emph{Proceedings of the 12th international conference on {Intelligent} {Virtual} {Agents}}} \emph{(\bibinfo{series}{{IVA}'12})}. \bibinfo{publisher}{Springer-Verlag}, \bibinfo{address}{Berlin, Heidelberg}, \bibinfo{pages}{289--302}.
\newblock
\showISBNx{978-3-642-33196-1}
\urldef\tempurl%
\url{https://doi.org/10.1007/978-3-642-33197-8_30}
\showDOI{\tempurl}


\end{thebibliography}

\end{document}